\documentclass[aps,prb,reprint,amsmath,amssymb,showpacs,floatfix]{revtex4-1}

\usepackage{graphicx,units,multirow}

\newcommand{\ket}[1]{\ensuremath{\vert{#1\rangle}}} 
\newcommand{\bra}[1]{\ensuremath{{\langle #1}\vert}}

\newcommand{\ketbra}[2]{\ensuremath{|{#1 \rangle}{\langle #2}|}}
\newcommand{\op}[1]{\hat{#1}}

\newcommand{\I}{\text{i}}
\newcommand{\E}{\text{e}}
\providecommand{\abs}[1]{\left\lvert#1\right\rvert}

\usepackage[colorlinks,allcolors=blue,hyperindex,breaklinks]{hyperref}

\begin{document}

\title{Assessing randomness with the aid of quantum state measurement}
\author{Mathew R.\ Coleman}
\author{Kaylin G.\ Ingalls}
\author{John T.\ Kavulich}
\author{Sawyer J.\ Kemmerly}
\author{Nicolas C.\ Salinas}
\author{Efrain Venegas Ramirez}
\author{Maximilian Schlosshauer}
\thanks{Author to whom correspondence should be addressed. Electronic mail: \texttt{schlossh@up.edu}}
\affiliation{Department of Physics, University of Portland, 5000 North Willamette Boulevard, Portland, Oregon 97203}

\begin{abstract}
 Randomness is a valuable resource in science, cryptography, engineering, and information technology. Quantum-mechanical sources of randomness are attractive because of the indeterminism of individual quantum processes. Here we consider the production of random bits from polarization measurements on photons. We first present a pedagogical discussion of how the quantum randomness inherent in such measurements is connected to quantum coherence, and how it can be quantified in terms of the quantum state and an associated entropy value known as min-entropy. We then explore these concepts by performing a series of single-photon experiments that are suitable for the undergraduate laboratory. We prepare photons in different nonentangled and entangled states, and measure these states tomographically. We use the information about the quantum state to determine, in terms of the min-entropy, the minimum amount of randomness produced from a given photon state by different bit-generating measurements. This is helpful in assessing the presence of quantum randomness and in ensuring the quality and security of the random-bit source.

\vspace{.1cm}

\noindent Journal reference: \emph{Am.\,J.\,Phys.\ }\textbf{88}, 238--246 (2020), DOI: \href{https://doi.org/10.1119/10.0000383}{\texttt{10.1119/10.0000383}} 
\end{abstract}

\maketitle

\section{Introduction}

Randomness plays an important role in science, engineering, technology, computing, and mathematics. For example, simulations of complex systems and phenomena often employ algorithms that rely on random numbers to account for effects that cannot, for reasons of computational cost or accuracy, be explicitly modeled.\cite{Motwani:1996:oo} In cryptography, secure communications necessitate the generation and distribution of random, secret keys for encrypting messages.\cite{Gisin:2002:ii} In fundamental quantum experiments, such as Bell tests\cite{Dehlinger:2002:tt} and delayed-choice experiments,\cite{Wheeler:1978:az} measurement settings must be chosen randomly to avoid loopholes. Computer networks, too, make use of randomness; for example, the Ethernet protocol assigns random wait times to minimize conflicts between nodes. Randomness is also an important resource in everyday applications such as gambling, lotteries, and computer games.

\begin{figure}
\includegraphics[scale=.9]{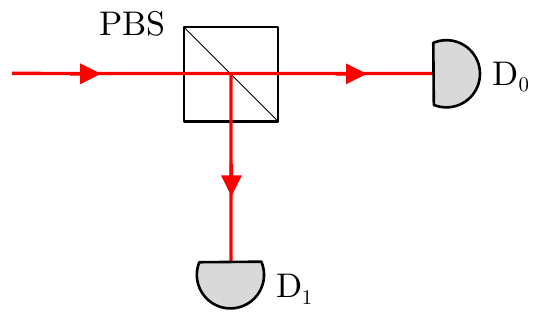}
\caption{\label{fig:concept}Basic principle of a branching-path quantum random number generator. A stream of photons, each prepared in the polarization state $\ket{\psi}=\frac{1}{\sqrt{2}} \left( \ket{H}+\E^{\I\phi}\ket{V}\right)$, is incident on a polarizing beam splitter (PBS), with detectors $D_0$ and $D_1$ placed at the outputs. This configuration realizes a polarization measurement in the horizontal--vertical ($HV$) basis. By identifying detection at $D_0$ with the bit 0 and detection at $D_1$ with 1, a random sequence of bits can be produced.}
\end{figure}

One common way of generating random numbers is to feed a starting value (the ``seed'') into a deterministic algorithm to produce a sequence of bits, typically with a uniform distribution. Such algorithmic methods are known as pseudorandom number generators (PRNGs; see Ref.~\onlinecite{Knuth:1997:ll} for a review). In contrast with PRNGs, physical random number generators  use a (fundamentally or practically) unpredictable physical process as a source of entropy for producing the bits. A particularly attractive source is provided by the indeterminism of individual quantum events, giving rise to quantum random number generators (QRNGs; see Ref.~\onlinecite{Herrero:2017:kk} for a review). The ease with which individual photons can now be produced, manipulated, and measured (even in undergraduate laboratories\cite{Dehlinger:2002:tt,Thorn:2004:za,Galvez:2005:uu,Dederick:2014:ll,Ashby:2016:pp,Beck:2012:az}) has put the focus on QRNGs that make use of the quantum properties of photons. In this paper, we will consider an implementation based on polarization measurements, shown in its basic form in Fig.~\ref{fig:concept}.

Randomness is commonly identified with a notion of unpredictability.\cite{Herrero:2017:kk} A lack of predictability might be due to insufficient information, or due to a fundamental indeterminism in nature (as described by quantum mechanics), or both. A crucial task associated with the realization of any bit-generating process is to quantify the amount of randomness produced by the source. Quantum mechanics offers unique opportunities in this regard, because its statistical character means that the relative frequencies of outcomes can be calculated if the quantum state is known, and because, as we will explain below, the amount of indeterminism in a measurement outcome can likewise be quantified.

In this paper, we explore how the measurement of quantum states can help us determine the amount of randomness in a process that produces bits from polarization measurements of photons. The purpose of our paper is twofold. First, it aims to provide an accessible, pedagogical discussion of the connections between quantum randomness, quantum coherence, and a widely used randomness measure known as the min-entropy.\cite{Cachin:1997:uu} Second, it experimentally illustrates these concepts and connections with the help of a single-photon apparatus that can be set up and operated by undergraduates. Our experiment uses a setup and components commonly found in other undergraduate teaching laboratories on single photons  (see Refs.~\onlinecite{Dehlinger:2002:tt,Thorn:2004:za,Galvez:2005:uu,Dederick:2014:ll,Ashby:2016:pp,Beck:2012:az} for examples), and thus it can be easily reproduced in such laboratories.

This paper is organized as follows. In Sec.~\ref{sec:randomness} we discuss different facets of randomness and their quantification, and describe how knowledge of the quantum state can be used to measure randomness. Here we make use of a result derived by Fiorentino \emph{et al.}\cite{Fiorentino:2006:lm,Fiorentino:2007:ll} that links the quantum state to the min-entropy.\cite{Cachin:1997:uu} In Sec.~\ref{sec:experiments} we experimentally illustrate these issues using a single-photon apparatus. We prepare photons in nonentangled and entangled polarization states and measure their states through a tomographic analysis. Using the quantum-state data together with the link between quantum states and min-entropy, we put a lower bound on the amount of randomness that could be extracted from bit-generating measurements of the kind shown in Fig.~\ref{fig:concept}.

\section{\label{sec:randomness}Theory}

\subsection{\label{sec:rand-as-unpr}Randomness as unpredictability}

We define the randomness of a source of bits in terms of the unpredictability of its output---i.e., how unexpected (surprising) a particular sequence of bits, and a particular bit in the sequence, are, relative to any information an observer may have about the bit-generation process and its output.\cite{Herrero:2017:kk}  
If, given such information, we cannot, on average, predict the next bit better than with a blind guess, then we conclude that we are dealing with a uniform random process (relative to the information we hold). We can refine this definition by introducing a measure of randomness (so that a source can be ``less'' or ``more'' random than another), defined in terms of the degree of predictability. If, on average, we can predict the outcome with better than a blind guess, then we have a certain amount of predictability, and hence only a certain amount of randomness. If we can predict the outcome with certainty, then we have perfect predictability and no randomness (we will make this quantification precise in Sec.~\ref{sec:min-entropy} below). 

For example, if we only have the bit strings produced by the source as our guide, we can look for patterns in the string. A pattern might be as simple as a bias toward 0 or 1 (nonuniformity). The bias reduces the degree of unpredictability, because we could use its observation to predict the next bit with, on average, better than a 50--50 chance. Other patterns, such as subtle periodicities, might only manifest themselves in very long sequences and be hard to distinguish. In general, we would need to search an infinite string for all possible patterns to rigorously rule out predictability, an impossible task.\cite{Li:2008:ii} (A formal definition of the presence of patterns is the Kolmogorov complexity\cite{Kolmogorov:1968:ii} of a sequence, but in its exact form it is uncomputable.\cite{Li:2008:ii}) We might also try to obtain information about the bit-generating process, for example, by learning about the particular way a coin is flipped. In this paper, we will use information about the quantum state to measure unpredictability.

\subsection{\label{sec:min-entropy}Min-entropy}

The notion of predictability can be quantified using measures of entropy. A widely used measure is the min-entropy.\cite{Cachin:1997:uu} It provides a worst-case bound on many other entropic measures and quantifies the effectiveness of any strategy that tries to guess, at first attempt, the most likely output of the source. An important meaning of min-entropy is that it gives the minimum number of uniform random bits that can be extracted from a given sequence,\cite{Chor:1988:ii} using a postprocessing technique known as randomness extraction.\cite{Nisan:1999:ll} 

For a binary process described by a random variable $X$, the min-entropy $H_\infty(X)$ (per bit) is defined as 
\begin{equation}\label{eq:minent3}
H_\infty(X) = -\log_2 \max \left( p_0, p_1\right),
\end{equation}
where $p_0$ and $p_i$ are the probabilities of the bits 0 and 1 (which may be interpreted as guessing probabilities for the next bit of the output). For a uniform probability distribution ($p_0=p_1=\frac{1}{2}$), the min-entropy assumes its maximum, $H_\infty = -\log_2 \frac{1}{2} = 1$. In this case, there is no better strategy than a blind guess, and we obtain a full random bit. If the result is predictable with certainty ($p_0=1$ or $p_1=1$), then $H_\infty = -\log_2 1 = 0$, representing a completely nonrandom process. If, more generally, $X$ takes values from the alphabet of all $2^N$ possible $N$-bit sequences, and if each sequence is equally likely, the min-entropy of $N$ bits generated by the source attains its maximum value of $H_\infty(X) = -\log_2 2^{-N} = N$.  An important practical problem is the estimation of the source's min-entropy.\cite{Herrero:2017:kk}  We will come back to this task below.

\subsection{Quantum randomness}

If we knew the seed and the algorithm of a PRNG, we could predict the output with certainty. This is analogous to the deterministic nature of classical physics: If we completely knew the laws and initial conditions, we could perfectly predict the outcome of, say, a coin toss. In such cases, the appearance of unpredictability is simply the result of a lack of information. Quantum mechanics, however, is fundamentally different, because it is not a deterministic theory. Knowing the quantum state, one can predict the relative frequencies of measurement outcomes, but in general one cannot, even in principle, know which particular outcome will occur in a given measurement. Crucially, this unpredictability is not something that could be overcome by gathering more information; rather, it is a property of nature rooted in the indeterministic character of the quantum measurement process. We shall refer to this property as \emph{quantum randomness}.\footnote{Here, we shall take the property of quantum randomness as given, but we note that it can be shown to be a consequence of any probabilitic theory that (i) violates Bell-type inequalities and (ii) does not permit superluminal signaling. See, e.g., L.~Masanes, A.~Acin, and N.~Gisin, ``General properties of nonsignaling theories,'' Phys. Rev. A, \textbf{73}, 012112 (2006), doi:10.1103/PhysRevA.73.012112.}  

Note that the notion of quantum randomness is distinct from the definition of randomness in terms of the degree of statistical unpredictability (especially concerning bias) introduced in Secs.~\ref{sec:rand-as-unpr} and \ref{sec:min-entropy}. In quantum mechanics, both notions are relevant. To see this, consider the pure quantum state $\cos\theta\ket{H}+\E^{\I\phi}\sin\theta\ket{V}$. The property of quantum randomness means that which particular outcome ($H$ or $V$) will occur in a polarization measurement in the horizontal--vertical ($HV$) basis is not predetermined by anything in nature. (From here on, we shall always consider the production of bits from such $HV$ measurements and therefore take the $HV$ basis as default.) Yet, if we know $\ket{\psi}$, then we can place a bet on whichever outcome is associated with the larger quantum probability ($p_H=\cos^2\theta$ or $p_V=\sin^2\theta$). since the resulting bit string will typically, in the long run, be biased toward the more likely outcome. It follows that in order to obtain a string that is both quantum-random and uniformly random, we need to consider $HV$ measurements on photons all prepared in the quantum state $\ket{\psi}=\frac{1}{\sqrt{2}} \left(\ket{H}+\E^{\I\phi} \ket{V}\right)$, such that $p_H=p_V$.

\subsection{\label{sec:rand-mixed-stat}Randomness for mixed states}

In general, a quantum system is described not by a pure state but by a mixed state, represented by a density operator $\op{\rho}$  (see Ref.~\onlinecite{Beck:2012:az} for an introduction to mixed states and density operators). A general mixed state can be written as $\op{\rho}=\sum_i p_i \ketbra{\psi_i}{\psi_i}$ with $0 < p_i \le 1$ and $\sum_i p_i=1$. Such a situation arises, for example, if there are fluctuations in the state preparation device, such that each preparation results in one of the pure states $\ket{\psi_i}$ with probability $p_i$, but we do not know in which. In this case, the probabilities $p_i$ are classical in the sense that they reflect our ignorance (rather than a fundamental indetermism); we say that the mixed state represents a classical ensemble of pure states. Mixed states also arise as the ``reduced'' states of a subsystem that is part of a larger system.\cite{Nielsen:2000:tt}  For example, for two photons prepared in the entangled Bell state $\ket{\Phi^+} = \frac{1}{\sqrt{2}} \left( \ket{H}\ket{H} + \ket{V}\ket{V}\right) $, the reduced state $\op{\rho}_r$ of one photon, obtained by a partial trace\cite{Nielsen:2000:tt,Dederick:2014:ll}  over the density operator $\op{\rho}= \ketbra{\Phi^+}{\Phi^+}$, is the mixed state $\op{\rho}_r=\frac{1}{2}\ketbra{H}{H}+\frac{1}{2}\ketbra{V}{V}$. The reduced state encapsulates the statistics of all possible measurements one can perform on this photon.

What distinguishes a pure state $\ket{\psi}=a\ket{H}+b\E^{\I\phi}\ket{V}$ (with $a$ and $b$ real, and $a^2+b^2=1$) from the mixed state $\op{\rho}=a^2\ketbra{H}{H}+b^2\ketbra{V}{V}$ is the presence of quantum coherence between the state components $\ket{H}$ and $\ket{V}$ in the pure state. Coherence between $\ket{H}$ and $\ket{V}$ represents the in-principle indistinguishability of $\ket{H}$ and $\ket{V}$ prior to the measurement, and thus implies the presence of quantum randomness in an $HV$ measurement. In the $HV$ basis, the matrix representation of the density operator $\op{\rho}=\ketbra{\psi}{\psi}$ associated with the pure state $\ket{\psi}$ is
\begin{equation}\label{eq:csvhcjh}
\rho = \begin{pmatrix}
\bra{H}\op{\rho}\ket{H} & \bra{H}\op{\rho}\ket{V} \\
\bra{V}\op{\rho}\ket{H} & \bra{V}\op{\rho}\ket{V} \end{pmatrix}=
\begin{pmatrix}
a^2 & ab\,\E^{-\I\phi}\\
ab\,\E^{\I\phi} & b^2 \end{pmatrix}.
\end{equation}
The magnitude $ab$ of the off-diagonal terms represents the amount of coherence between $\ket{H}$ and $\ket{V}$. It gets smaller as the amplitudes $a$ and $b$ become more different. The off-diagonal terms (and thus quantum randomness) are maximized when $a^2=b^2=\frac{1}{2}$, in which case we also have a uniform random process, since the probabilities of $H$ and $V$ are equalized. By comparison, the density matrix for the mixed state $\op{\rho}=a^2\ketbra{H}{H}+b^2\ketbra{V}{V}$ is
\begin{equation}\label{eq:vdlkndv2}
\rho = 
\begin{pmatrix}
a^2 & 0\\
0 & b^2 \end{pmatrix}.
\end{equation}
This state gives the same probabilities for $HV$ measurements as the state~\eqref{eq:csvhcjh}, but now the off-diagonal elements are zero, i.e., there is no coherence between $\ket{H}$ and $\ket{V}$, and therefore no guaranteed quantum randomness in an $HV$ measurement.\footnote{We emphasize here that the condition of coherence between $\ket{H}$ and $\ket{V}$ is not the same as the purity of the quantum state. The state $\op{\rho}=\ketbra{H}{H}$ is pure but trivially has no coherence between $\ket{H}$ and $\ket{V}$ (the off-diagonal elements in the $HV$ basis are zero).} In general, any quantum state $\op{\rho}$ can be written as\cite{Nielsen:2000:tt} (the $HV$ basis is again implicit) 
\begin{equation}\label{eq:lkdvbjb1}
\rho =  
\begin{pmatrix}
A & C\E^{\I\phi} \\
C\E^{-\I\phi}  & B \end{pmatrix},
\end{equation}
with $A$, $B$, and $C$ real and nonnegative, $A+B=1$, and $C \le \sqrt{AB}$, where the equality holds for a pure state [compare Eq.~\eqref{eq:csvhcjh}]. 

As mentioned, a state such as~\eqref{eq:vdlkndv2} might represent a classical ensemble or the reduced state. Let us illustrate these two cases using the example of generation of random keys in cryptography. For the case in which the mixed state is a classical ensemble, we might imagine an adversary, Eve, who prepares a stream of photons, each in either the state $\ket{H}$ or the state $\ket{V}$, and then feeds the photons to Alice. To Eve, the collection of photons is in a pure state (since she knows the polarization of each photon), but Alice would need to assign the mixed state $\op{\rho}_r=p_H\ketbra{H}{H}+p_V\ketbra{V}{V}$ (since each photon is in a pure state but she does not know in which). To Alice, the bit string resulting from her $HV$ measurements will appear random, but there is no quantum randomness involved in her measurements---Eve will be able to perfectly predict each bit Alice produces. Alice's key would not be private.

For the case of a reduced state, consider the scenario in which Alice performs $HV$ measurements on one photon in the entangled state $\ket{\Phi^+} = \frac{1}{\sqrt{2}} \left( \ket{H}\ket{H} + \ket{V}\ket{V}\right)$ (we explore this scenario experimentally in Sec.~\ref{sec:meas-one-phot}). The quantum correlations represented by $\ket{\Phi^+}$ imply that no coherence between $\ket{H}$ and $\ket{V}$ can be observed locally by measuring just one of the photons in the pair. This is so because one could, in principle, fix the state ($\ket{H}$ or $\ket{V}$) of one photon in the pair by measuring the other photon. Whether such a measurement is actually carried out is irrelevant for this loss of local coherence. If Eve possesses the other photon in the entangled pair and does measure it in the $HV$ basis, then the outcomes of the $HV$ measurements performed by Alice and Eve on their respective photons will be perfectly correlated, so Eve would be able to learn Alice's sequence. (If she measures before Alice, we have effectively the scenario described in the previous paragraph.) Because the statistics of any measurements Alice could undertake on her photon cannot be influenced by Eve's measurements on the other photon, the density matrix of Alice's photons must reflect the mere \emph{possibility} that Eve's measurements could have taken place. This implies a description by an incoherent mixture of $\ket{H}$ and $\ket{V}$, $\op{\rho}_r=\frac{1}{2}\ketbra{H}{H}+\frac{1}{2}\ketbra{V}{V}$, which is Alice's reduced state. Again, there is no guaranteed quantum randomness for Alice's measurements.

In summary, the magnitude of the off-diagonal elements of the density matrix expressed in the $HV$ basis indicates the amount of randomness we can generate through $HV$ measurements. If the off-diagonal elements are zero [as in Eq.~\eqref{eq:vdlkndv2}], then there is no guarantee that the output is not predetermined (in the sense that it could be predicted with certainty). If the off-diagonal elements are nonzero, then there is at least some degree of quantum randomness in the output, and thus we can generate fresh randomness even if the state is fully known. We can increase this randomness by increasing coherence between $\ket{H}$ and $\ket{V}$, and by making the probabilities of the outcomes $H$ and $V$ more similar (which also decreases bias).

\subsection{\label{sec:min-entropy-bound}Min-entropy bound from the quantum state}

As discussed in Sec.~\ref{sec:min-entropy}, the min-entropy of a source measures the degree of randomness (unpredictability) of its output. Then, in the previous Sec.~\ref{sec:rand-mixed-stat}, we saw how we can use knowledge of the quantum state of the photon to quantify the amount of randomness generated by an $HV$ measurement. We will now connect these concepts by describing how we can use knowledge of the quantum state to estimate the min-entropy. 

\begin{figure}
\includegraphics[scale=0.55]{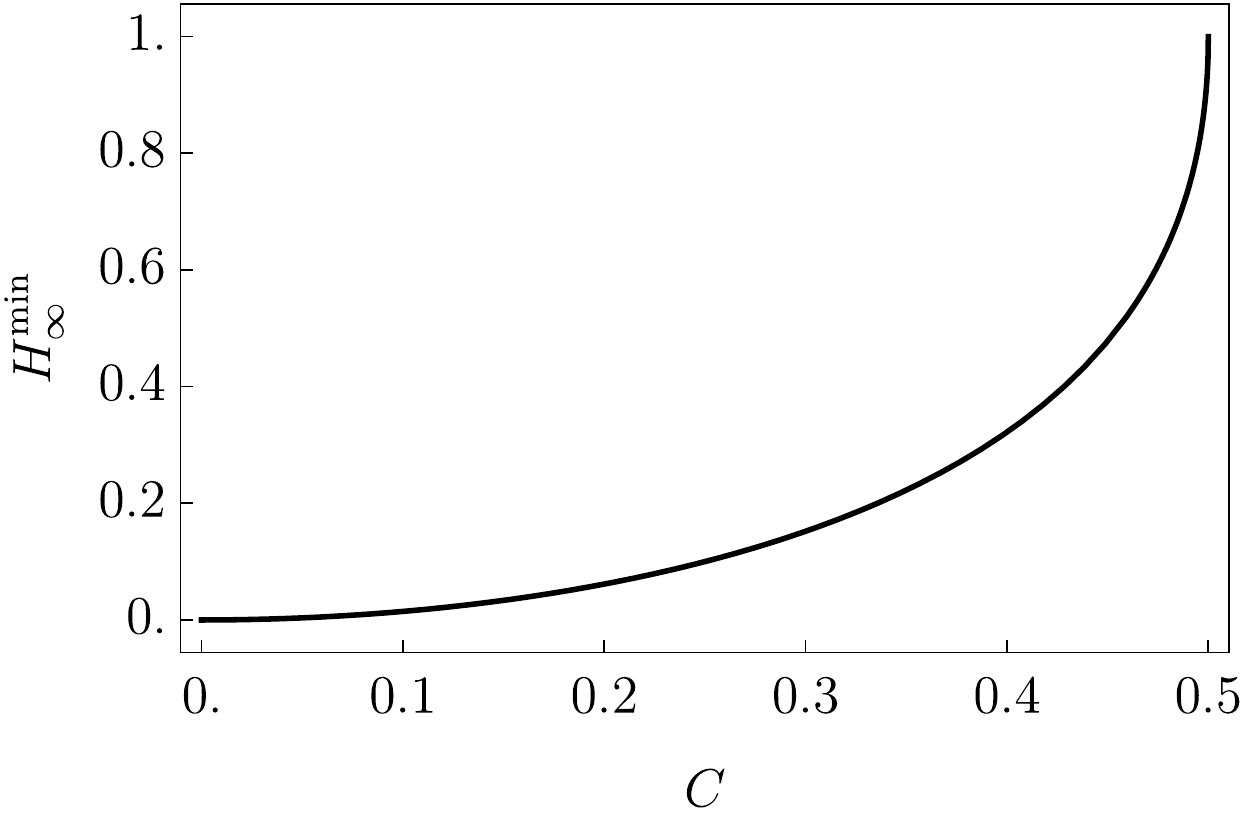}
\caption{\label{fig:minent}Lower bound $H^\text{min}_\infty$ [see Eq.~\eqref{eq:fio}] on the min-entropy per $HV$ measurement when the state $\op{\rho}$ is known, shown as a function of the size $C$ of the off-diagonal elements of the density matrix expressed in the $HV$ basis.}
\end{figure}

For a pure state $\ket{\psi}=a\ket{H}+b\E^{\I\phi}\ket{V}$, there is maximum quantum randomness in the outcome of the $HV$ measurement, and we can simply identify the probabilities $p_0$ and $p_1$ in the general expression~\eqref{eq:minent3} for the min-entropy (now per $HV$ measurement) with the quantum probabilities $a^2$ and $b^2$ specified by $\ket{\psi}$,
\begin{equation}\label{eq:minentq}
H_\infty(\ketbra{\psi}{\psi}) = -\log_2 \max \left( a^2, b^2\right).
\end{equation}
On the other hand, for the mixed state $\op{\rho}=a^2\ketbra{H}{H}+b^2\ketbra{V}{V}$ we cannot guarantee that the measurement produces fresh randomness, and hence there is no guaranteed unpredictability. Thus, the min-entropy might be zero and we cannot do better than provide a lower bound of $H_\infty(\op{\rho}) \ge 0$. For a general photon state~\eqref{eq:lkdvbjb1}, since the degree of quantum randomness for $HV$ measurements is given by the size of the off-diagonal elements, one can use this size to put a lower bound on the min-entropy of the source. Fiorentino \emph{et al.}\cite{Fiorentino:2006:lm,Fiorentino:2007:ll} showed that for photons in a known state $\op{\rho}$, the min-entropy $H_\infty(\op{\rho})$ per $HV$ measurement is no less than
\begin{equation}\label{eq:fio}
H^\text{min}_\infty(\op{\rho}) = -\log_2\left(\frac{1+\sqrt{1-4C^2}}{2} \right),
\end{equation}
where $C$ is the magnitude of the off-diagonal elements of $\op{\rho}$ expressed in the $HV$ basis [see Eq.~\eqref{eq:lkdvbjb1}]. A plot of $H^\text{min}_\infty$ as a function of $C$ is shown in Fig.~\ref{fig:minent}. (Note that we must also have $H_\infty(\op{\rho}) \le 1$, since one cannot generate more than one random bit per measurement.) 

Let us check Eq.~\eqref{eq:fio} for some important cases. For the mixed state $\op{\rho}=a^2\ketbra{H}{H}+b^2\ketbra{V}{V}$, we have $C=0$ [see Eq.~\eqref{eq:vdlkndv2}] and therefore $H^\text{min}_\infty(\op{\rho}) =  -\log_2 1 = 0$, as desired. For the pure state $\ket{\psi}=\frac{1}{\sqrt{2}} \left(\ket{H}+\E^{\I \phi}\ket{V}\right)$, we have $C=\frac{1}{2}$ and $H^\text{min}_\infty(\op{\rho}) = 1$, and thus we reach the maximum $H_\infty(\op{\rho})=1$. For a pure state $\ket{\psi}=a\ket{H}+b\E^{\I\phi}\ket{V}$, $C=ab=a\sqrt{1-a^2}$ and the parenthetical expression on the right-hand side of Eq.~\eqref{eq:fio} evaluates to $\frac{1}{2} + \abs{\frac{1}{2}-a^2}$. If $a^2 < \frac{1}{2}$, then this expression is equal to $1-a^2=b^2$, which is also $\max(a^2,b^2)$. Otherwise, it is equal to $a^2$, which again is $\max(a^2,b^2)$.  Thus, Eq.~\eqref{eq:fio} is equal to $-\log_2 \max(a^2,b^2)$ and we recover Eq.~\eqref{eq:minentq}. This shows that for a pure state, the min-entropy is equal to $H^\text{min}_\infty(\op{\rho})$.

Equation~\eqref{eq:fio} gives the min-entropy bound when the quantum state is known. This is useful when we want to use quantum state measurement to estimate the randomness produced by polarization measurements (as we do in Sec.~\ref{sec:experiments}). In particular, because $H^\text{min}_\infty(\op{\rho})$ is nonzero if and only if there is coherence between $\ket{H}$ and $\ket{V}$, we can use it to ensure quantum randomness in the outcomes of $HV$ measurements. Equation~\eqref{eq:fio} is also useful in the adversarial scenarios of Sec.~\ref{sec:rand-mixed-stat}, where Eve might have prepared the quantum states that Alice is measuring, or when Alice's photons are quantum-correlated with Eve's. Then the quantum randomness guaranteed by a nonzero min-entropy bound~\eqref{eq:fio} ensures that Alice's measurements will produce fresh randomness that will allow her to extract a certain minimum amount (given by $H^\text{min}_\infty$) of uniform, private random bits uncorrelated with any of Eve's information.\cite{Fiorentino:2006:lm,Fiorentino:2007:ll} That is, Eq.~\eqref{eq:fio} gives the maximum amount of information (per bit) that Eve can obtain about Alice's sequence. We experimentally study such an adversarial scenario in Sec.~\ref{sec:meas-one-phot}.

\section{\label{sec:experiments}Experiments}

We now turn to our experiment, in which we consider different scenarios for the production of bits from $HV$ polarization measurements and then quantify the randomness of the process in each scenario. To do so, for each scenario we measure the quantum state, and then apply Eq.~\eqref{eq:fio} to determine a lower bound on the min-entropy of the bit-generating process. 

\subsection{Experimental apparatus}

\begin{figure}
\includegraphics[scale=.9]{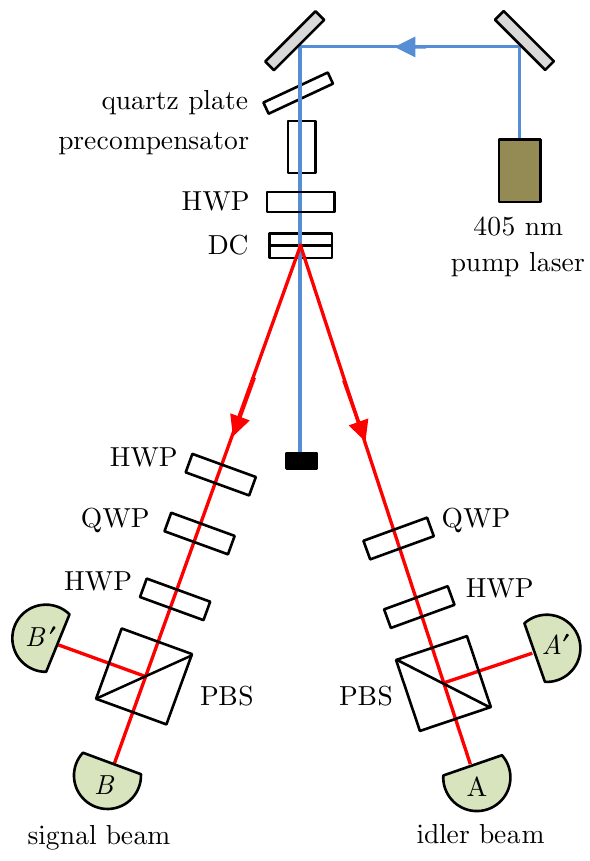}
\caption{\label{fig:setup}Schematic of the experimental setup. Here DC denotes the down-conversion crystals, HWP denotes a half-wave plate, QWP denotes a quarter-wave plate, and PBS denotes a polarizing beam splitter. The outputs from the beam splitters are fed into fiber-coupled lenses and detected by single-photon counting modules.}
\end{figure}

Our experimental apparatus, shown in Fig.~\ref{fig:setup}, is capable of producing polarization-entangled photon pairs and is similar to the undergraduate-compatible photonic setups that have been described in this journal;\cite{Dehlinger:2002:tt,Thorn:2004:za,Galvez:2005:uu,Dederick:2014:ll,Ashby:2016:pp} see also Ref.~\onlinecite{Beck:2012:az} for details. A 405-nm, 150-mW diode pump laser is incident on a pair of closely stacked, 0.5-mm-thick beta-barium borate crystals cut for type-I spontaneous parametric downconversion.\cite{Kwiat:1999:oo,Thorn:2004:za,Beck:2012:az}  The optic axes of the two crystals are oriented at right angles to each other, such that one of the crystals produces pairs of horizontally polarized 810-nm photons while the other produces vertically polarized pairs. Emitted photons make an angle of about $3^\circ$ with the pump beam. We refer to the two photon beams as signal and idler. Using a half-wave plate, the pump polarization can be rotated to pump either just one downconversion crystal (to produce a nonentangled state $\ket{H}\ket{H}$), or both crystals to produce an entangled state resembling $\ket{\Phi^+} = \frac{1}{\sqrt{2}} \left( \ket{H}\ket{H} + \E^{\I \phi} \ket{V}\ket{V}\right)$. The phase $\phi$ can be adjusted by rotating an X-cut, $\unit[10 \times 10 \times 0.5]{mm}$ quartz plate placed upstream from the downconversion crystal. To enhance entanglement, we precompensate for the walk-off of the orthogonal polarization components inside the downconversion crystal by inserting a $\unit[5 \times 5 \times 5.58]{mm}$ quartz crystal.

The signal and idler photons are subjected to polarization analyzers, each consisting of a quarter-wave plate, a half-wave plate, and a polarizing beam splitter. By turning the wave plates to appropriate settings, polarization measurements in three different bases can be realized for each photon: the $HV$ basis, the diagonal basis defined by $\ket{D}=\frac{1}{\sqrt{2}} \left(\ket{H}+\ket{V}\right)$ and $\ket{A}=\frac{1}{\sqrt{2}} \left(\ket{H}-\ket{V}\right)$, and the circular basis defined by $\ket{R}= \frac{1}{\sqrt{2}} \left(\ket{H}+\I\ket{V}\right)$ and $\ket{L}=\frac{1}{\sqrt{2}} \left(\ket{H}-\I\ket{V}\right)$. Photons are captured by converging lenses coupled to multimode fiber-optic cables and transmitted to single-photon counting modules based on silicon avalanche photodiodes (the detection efficiency is about 30\% at \unit[810]{nm}). Ambient photons are removed by 780-nm long-pass filters placed at the inputs of the counting modules. To ensure detection of single photons, we measure signal and idler photons in coincidence within a time window of  \unit[7--8]{ns}.\cite{Lord:noyear:uu} Coincidences are processed by a field-programmable gate array implemented on an Altera DE2 board,\cite{Lord:noyear:uu} which transmits data to a PC running Lab\textsc{view} software.\cite{Beck:website} 

In our experiment, we do not actually read out a bit string produced from the outcomes of individual polarization measurements, because our equipment cannot time-stamp individual photon events; instead, it accumulates photon counts over a preset time interval. Nor does our equipment allow for making the counting interval short enough such that predominantly no photons or just one photon are registered, which could be used to define bits from individual outcomes.\cite{Branning:2010:oo} To be sure, we could instead produce bits from, e.g., the parity (even or odd) of photon counts in each counting interval. We do not, however, pursue this approach here, because it would not allow for the randomness of the resulting string to be directly related to the min-entropy estimation for individual measurement events, and because the focus of our paper is on the use of quantum state measurement for assessing randomness.

\subsection{\label{sec:quant-state-tomogr}Quantum state tomography}

\begin{figure*}
\includegraphics[scale=.4]{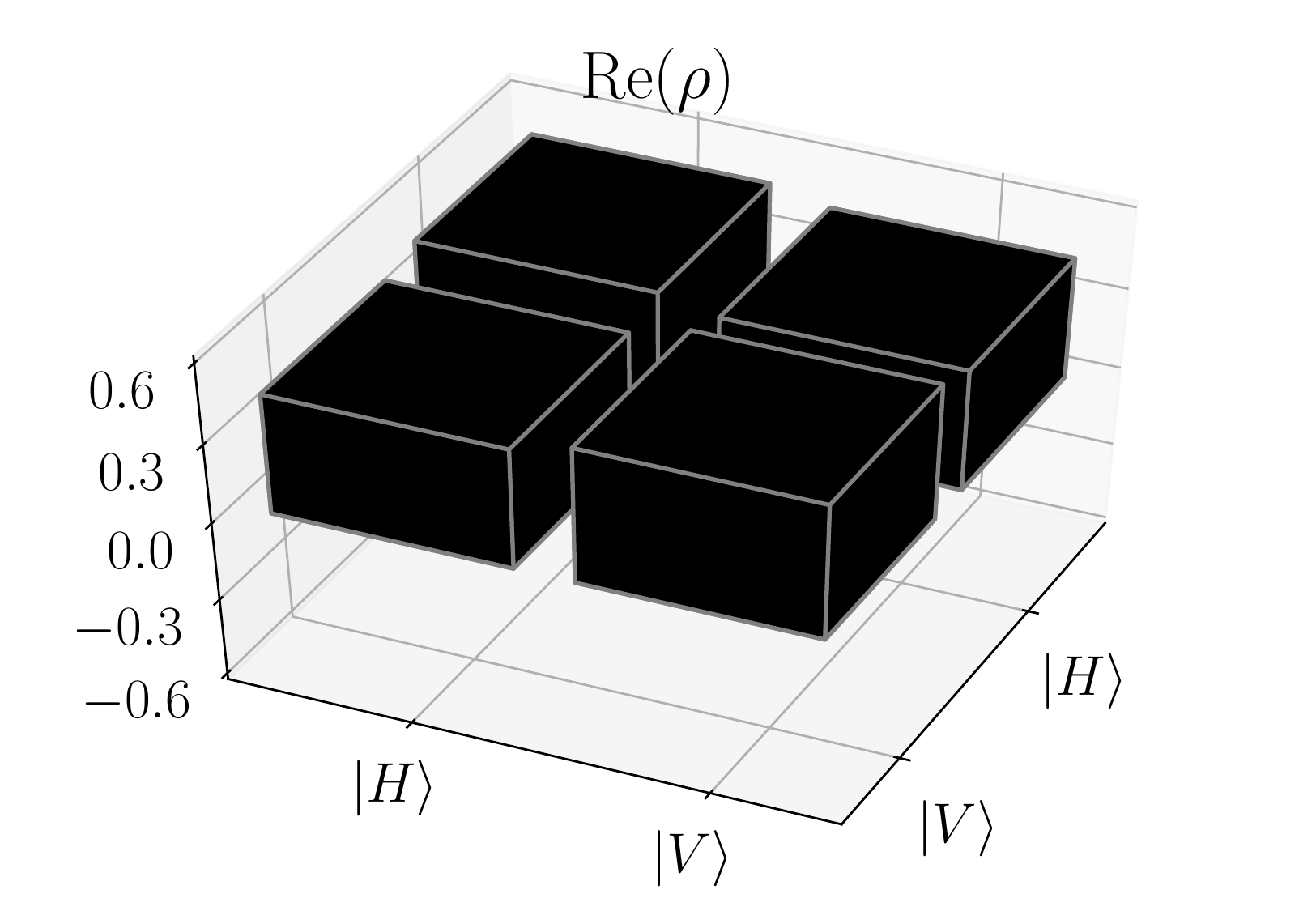} \includegraphics[scale=.4]{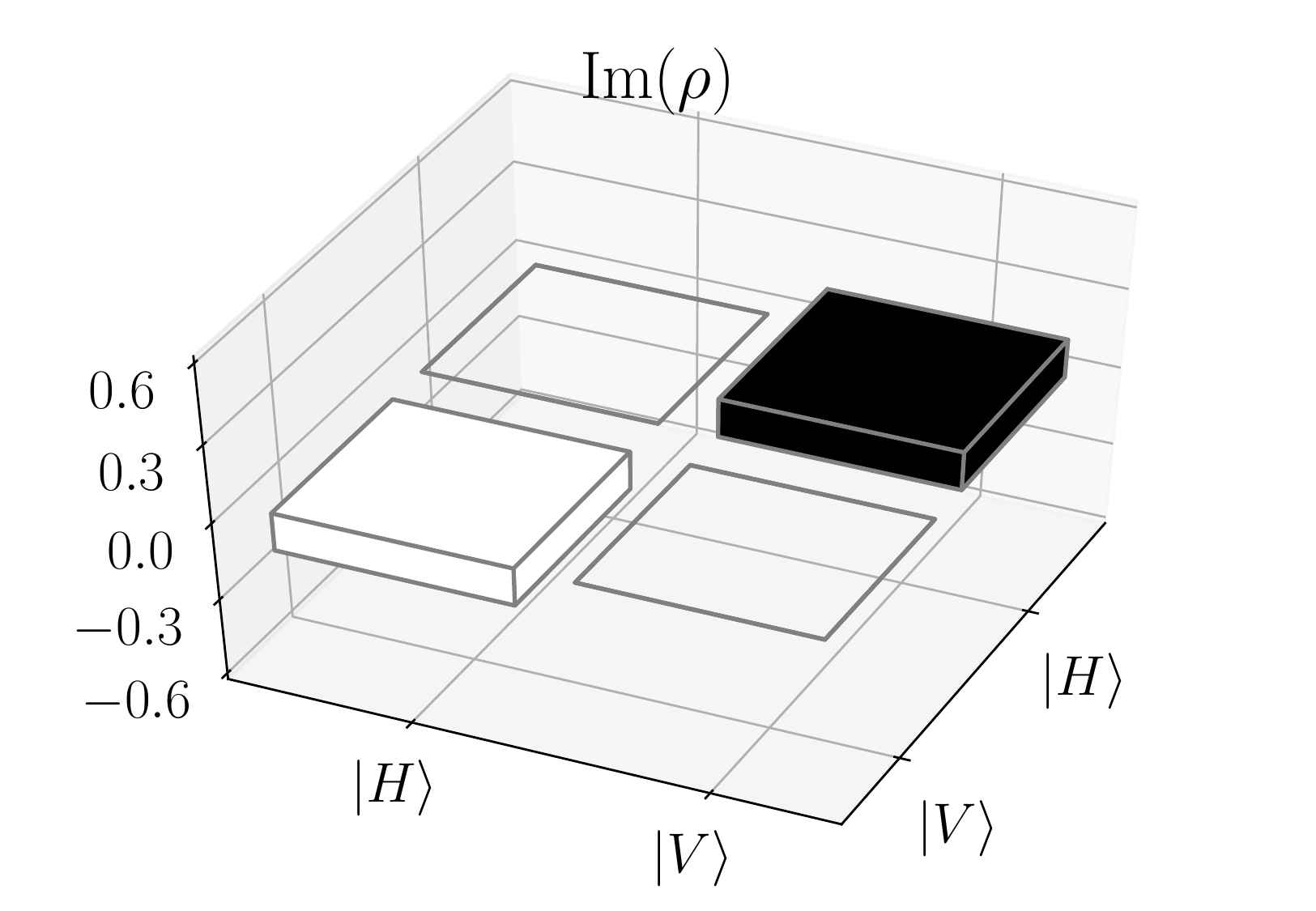} 
\caption{\label{fig:nonent}Real (left) and imaginary (right) parts of the density matrix for the signal photon prepared in a superposition of horizontal and vertical polarizations. Black bars represent positive values, white bars negative values.}
\end{figure*}

In photonic quantum state tomography, the density matrix of an ensemble of photons is reconstructed from a series of polarization measurements in different bases. We refer the reader to Refs.~\onlinecite{Altepeter:2005:ll,Beck:2012:az,Dederick:2014:ll} for introductions to the subject, and here just briefly review the main idea. The density operator for the photon can always be written as
\begin{equation}\label{eq:dm2}
\op{\rho} = \frac{1}{2} \sum_{i=0}^3 S_i \op{\sigma}_i,
\end{equation}
where the $\op{\sigma}_i$ are the Pauli matrices, and the $S_i = \langle \sigma_i \otimes \sigma_j \rangle$ are expectation values (in this context known as Stokes parameters) that can be measured by performing projective measurements on the photons in the $HV$ basis, the diagonal basis ($DA$), and the circular basis ($RL$).\cite{Altepeter:2005:ll,Beck:2012:az,Dederick:2014:ll} The half-wave and quarter-wave plate settings are $(0^\circ,0^\circ)$ for a measurement in the $HV$ basis, $(22.5^\circ,45^\circ)$ for the $DA$ basis, and $(0,45^\circ)$ for the $RL$ basis.\cite{Beck:2012:az} From these three measurement settings and the statistics of the coincidence counts measured at each setting, the Stokes parameters can be estimated and the density matrix can be reconstructed.

This method readily generalizes to the measurement of the  $4\times 4$ density operator of a pair of photons.  This density operator can be written as
\begin{equation}\label{eq:dm3}
\op{\rho} = \frac{1}{4} \sum_{i,j=0}^3 S_{ij} \left(\op{\sigma}_i \otimes \op{\sigma}_j\right).
\end{equation}
Here, the $S_{ij} = \langle \sigma_i \otimes \sigma_j \rangle$ are the (two-photon) Stokes parameters, which we can estimate by performing polarization measurements in the three bases on both photons.

\subsection{\label{sec:bit-generation-from}Bit generation from measurements of photons in a nonentangled state}

First, we create photon pairs in the nonentangled state $\ket{H}\ket{H}$ by pumping only one of the downconversion crystals. We use a half-wave plate in the signal beam to prepare the signal photon in a state close to the diagonal state $\ket{D}=\frac{1}{\sqrt{2}} \left(\ket{H}+\ket{V}\right)$ and remove all other wave plates in the beam. On average, half of the photons will be registered at the $B$ output of the beam splitter and half at $B'$ (compare Fig.~\ref{fig:setup}). This realizes an $HV$ measurement and thus represents the standard branching-path method for bit generation shown in Fig.~\ref{fig:concept}. 

To estimate the min-entropy of such a random process, we perform quantum state tomography on the signal photon by inserting a half-wave plate and quarter-wave plate into the signal beam and carrying out measurements in the $HV$, diagonal, and circular bases as described in Sec.~\ref{sec:quant-state-tomogr}. The resulting density matrix $\rho$ is
\begin{equation}
\rho = \begin{pmatrix} 0.493 & 0.449 +0.144 \I \\
0.449 -0.144\I & 0.507 \end{pmatrix},
\end{equation}
which is shown in Fig.~\ref{fig:nonent}. 

First, we assess the similarity of this state to the diagonal state $\op{\rho}_D=\ketbra{D}{D}$ by calculating the fidelity\cite{Jozsa:1994:oo} $F(\rho,\rho_D)=\left(\text{Tr} \sqrt{ \sqrt{\rho}\rho_D\sqrt{\rho}}\right)^2$ for the two density matrices. We find $F=0.974$, which shows that the prepared state is indeed well described by $\ket{D}$. The diagonal elements (i.e., the probabilities for $H$ and $V$) are similar in size ($\rho_{11}=p_H=0.493$ and $\rho_{22}=p_V=0.507$), giving near-uniformity of the bit sequence in the long run. The size of the off-diagonal elements is $C=\abs{\rho_{12}}=\abs{\rho_{21}}=0.472$, which is close to the maximum value of $C_\text{max}=\sqrt{\rho_{11}\rho_{22}}=0.5$ that would be attained for a pure state $\ket{\psi}=\sqrt{\rho_{11}}\ket{H}+\E^{\I\phi}\sqrt{\rho_{22}}\ket{V}$ [compare Eq.~\eqref{eq:lkdvbjb1}]. This indicates a large amount of coherence between $\ket{H}$ and $\ket{V}$, and thus a large amount of quantum randomness in the outcomes of $HV$ measurements. We calculate the min-entropy bound~\eqref{eq:fio} from the size of the off-diagonal elements and obtain $H^\text{min}_\infty = 0.589$. Recall that the min-entropy gives the number of uniform random bits (per bit) that can be extracted from the raw bit string.\cite{Chor:1988:ii} Thus, the value of the bound signals that from $N$ bits obtained from $HV$ measurements, one could extract a uniform random string containing at least $0.589 N$ bits.

\begin{figure*}
\includegraphics[scale=.40]{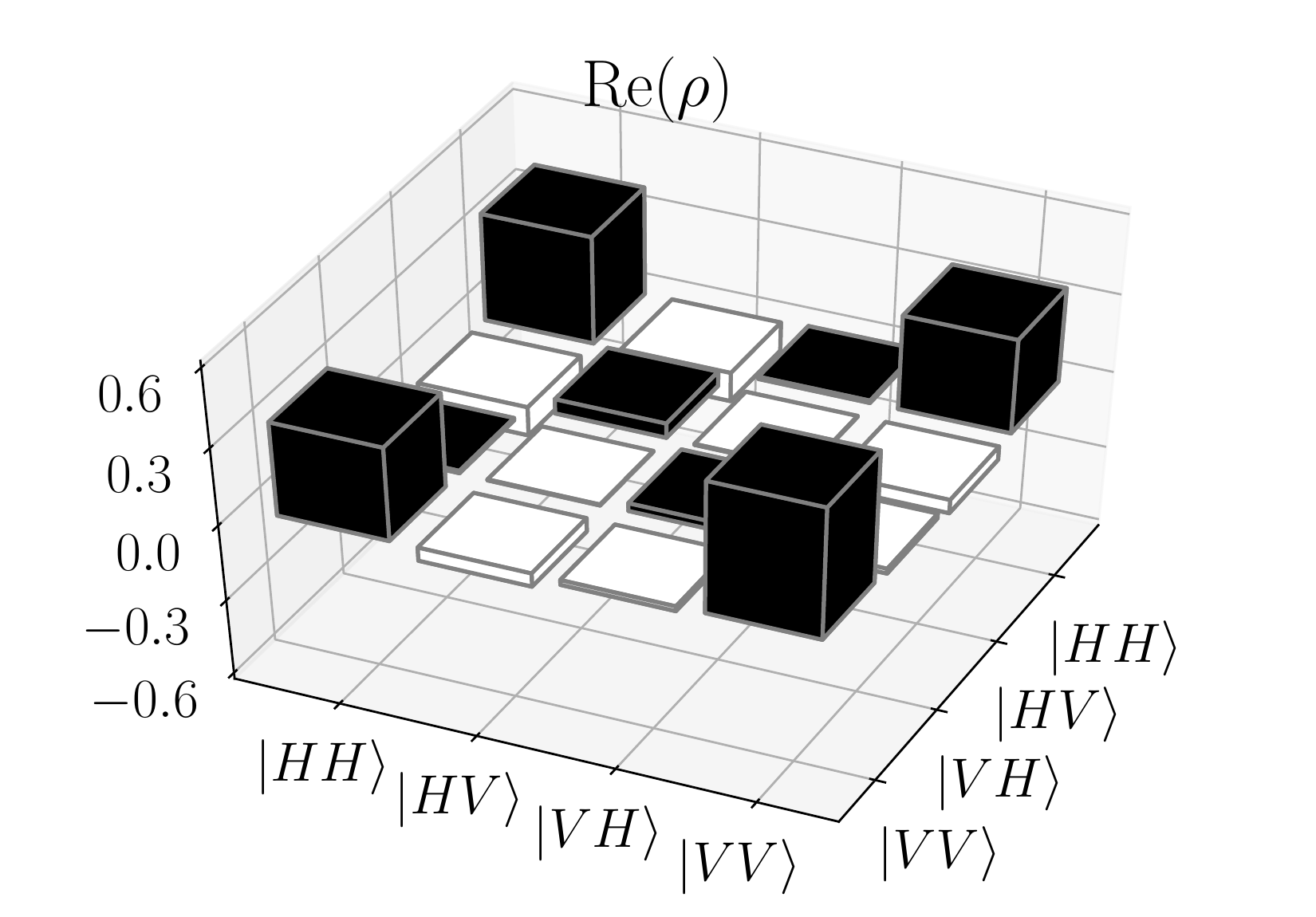} \includegraphics[scale=.40]{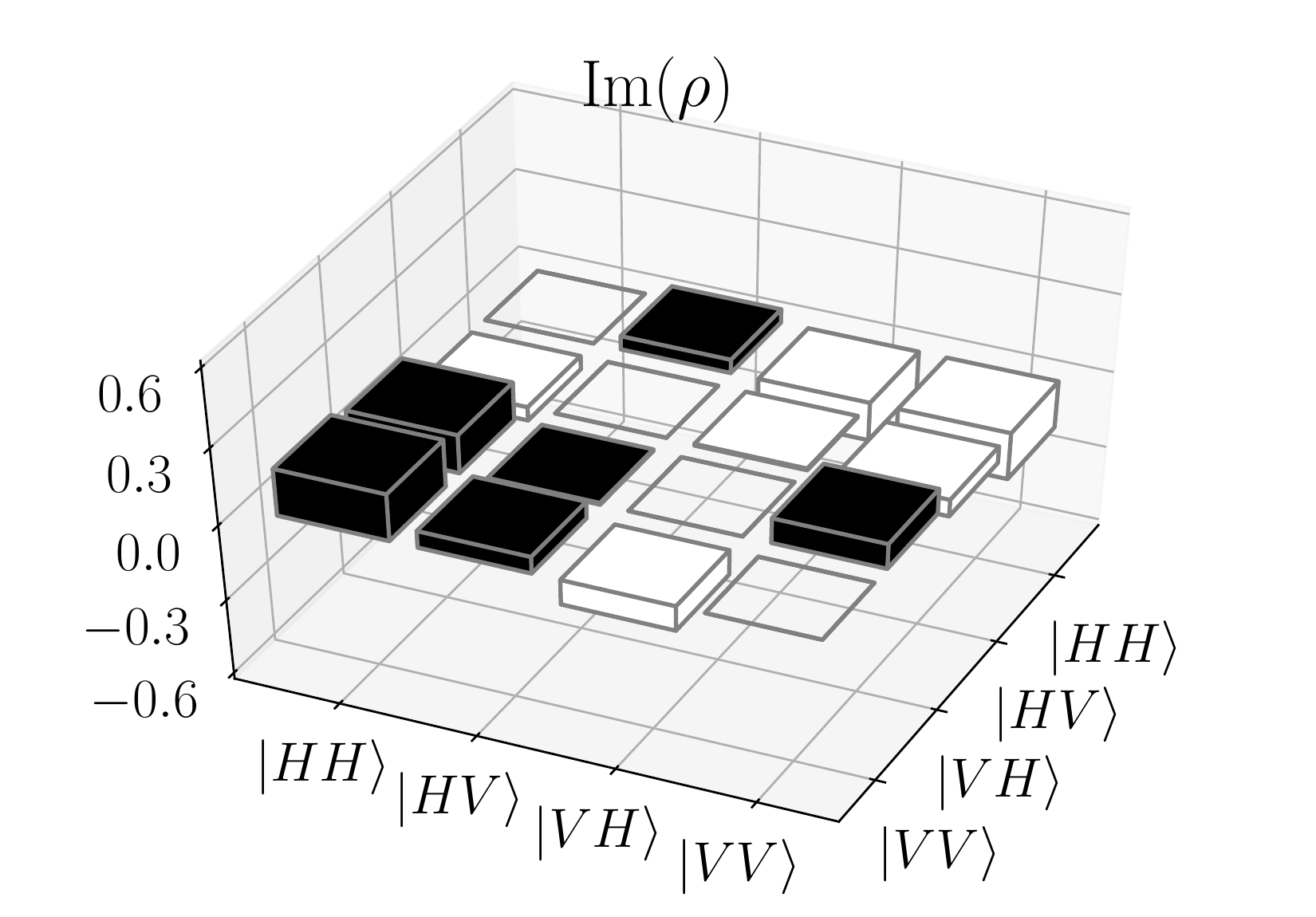} 
\caption{\label{fig:tomo4d}Real (left) and imaginary (right) parts of the density matrix reconstructed from quantum state tomography of the two-photon state. Black bars represent positive values, white bars negative values.}
\end{figure*}

\subsection{\label{sec:photon-pair-an}Bit generation from measurements of photon pairs in an entangled state}

\begin{figure*}
\includegraphics[scale=.40]{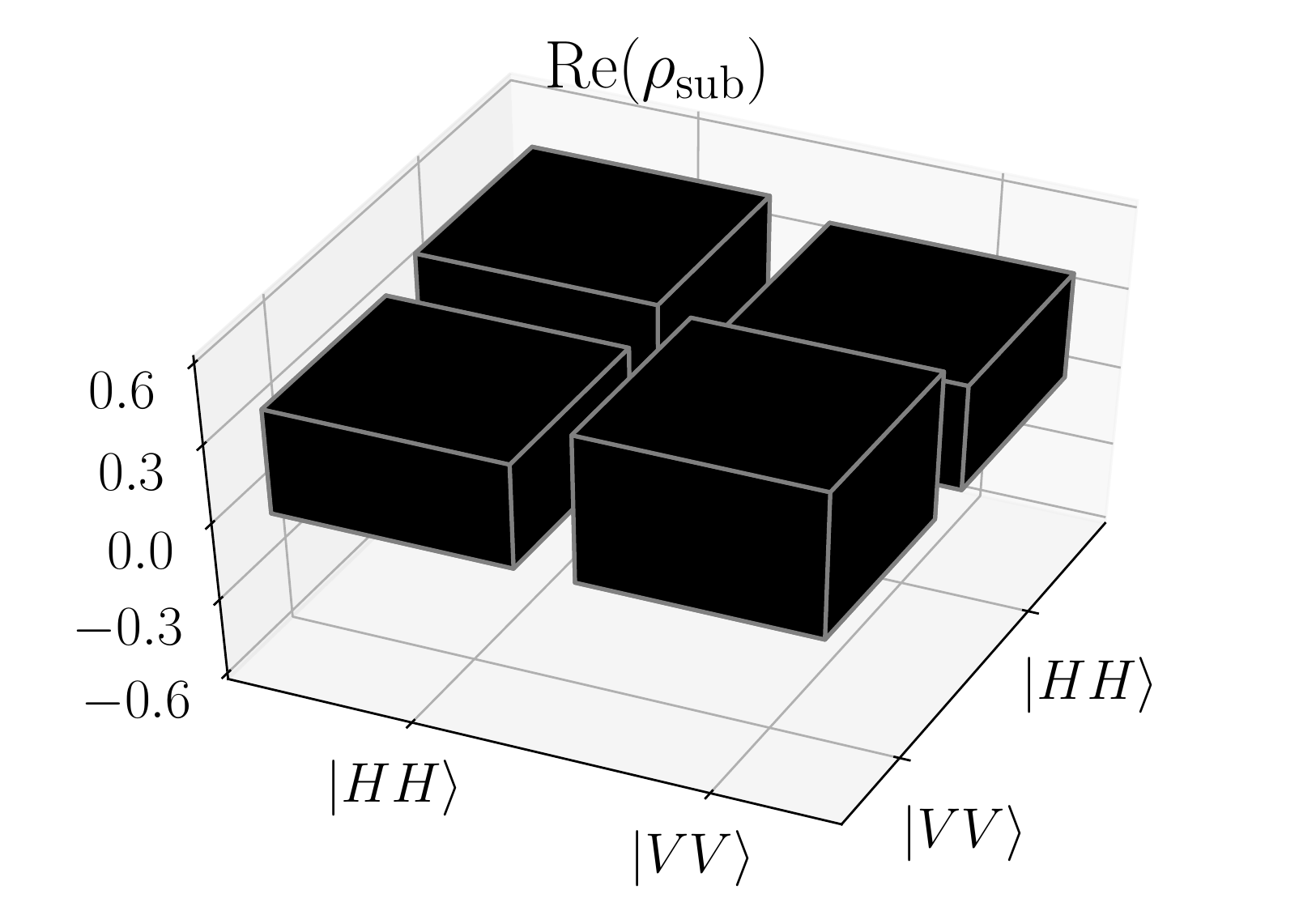} \includegraphics[scale=.40]{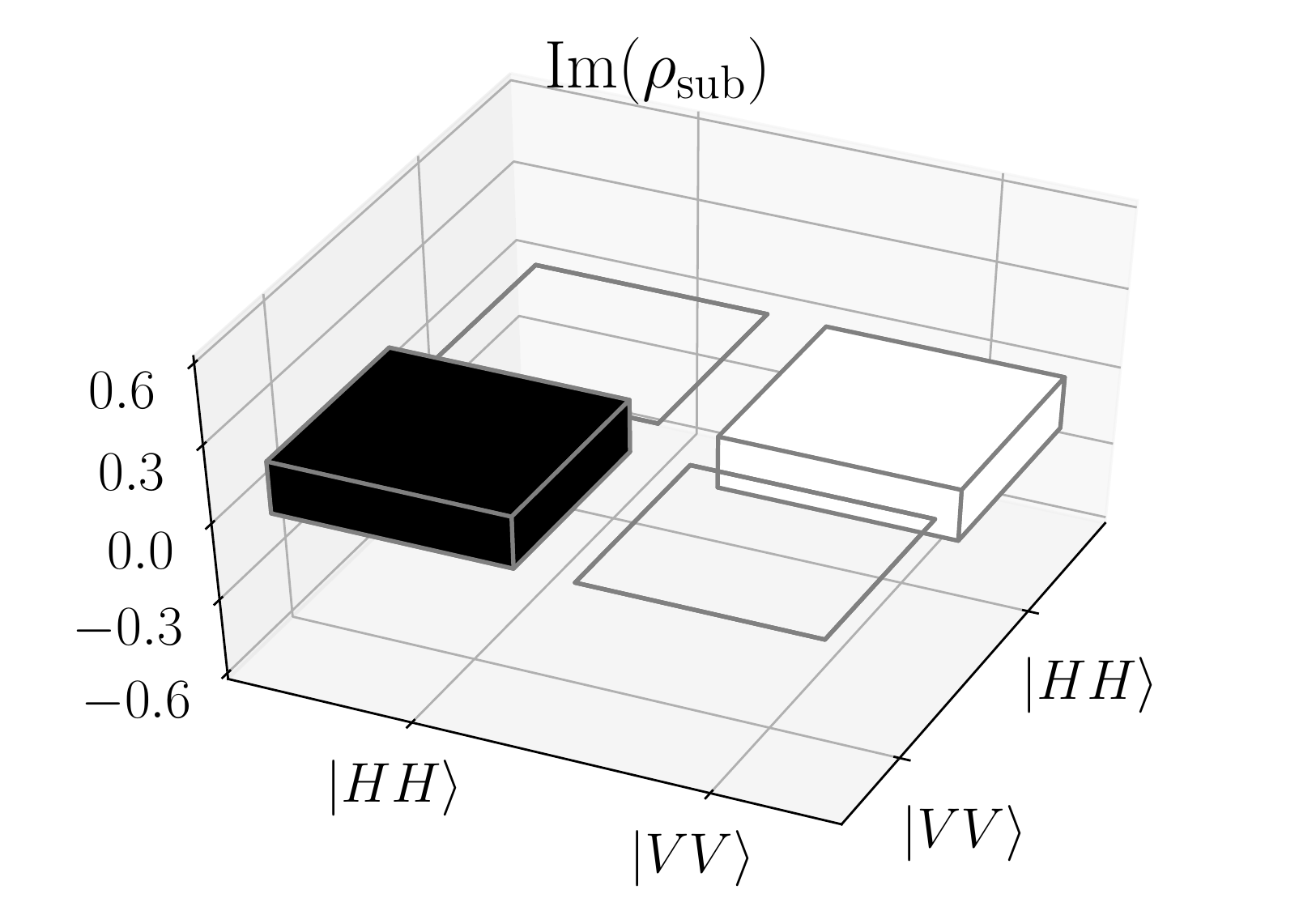} 
\caption{\label{fig:sub}Real (left) and imaginary (right) parts of the two-photon density matrix for the subspace spanned by the coincidence events $HH$ and $VV$. Black bars represent positive values, white bars negative values.}
\end{figure*}

\begin{figure*}
\includegraphics[scale=.40]{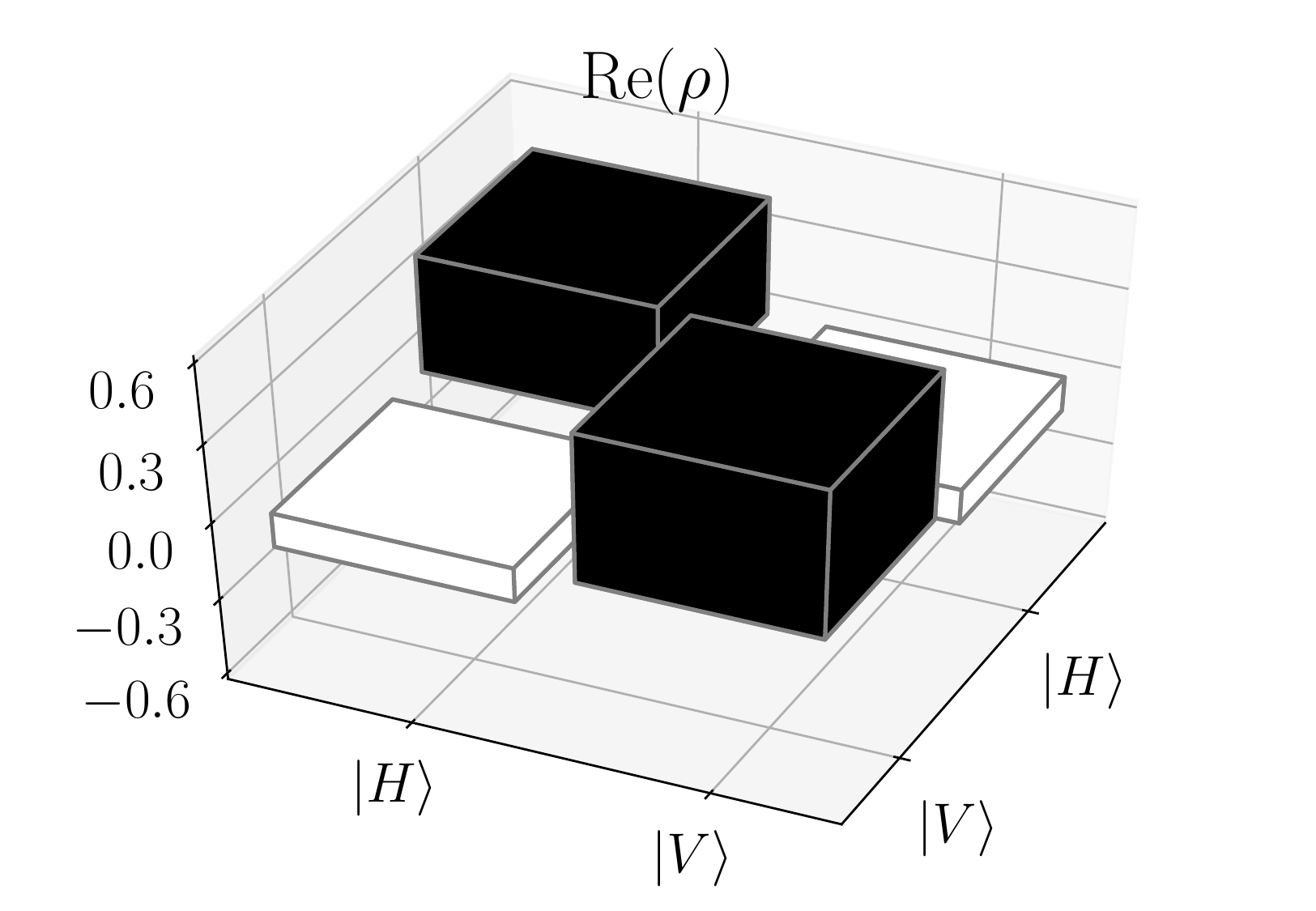} \includegraphics[scale=.40]{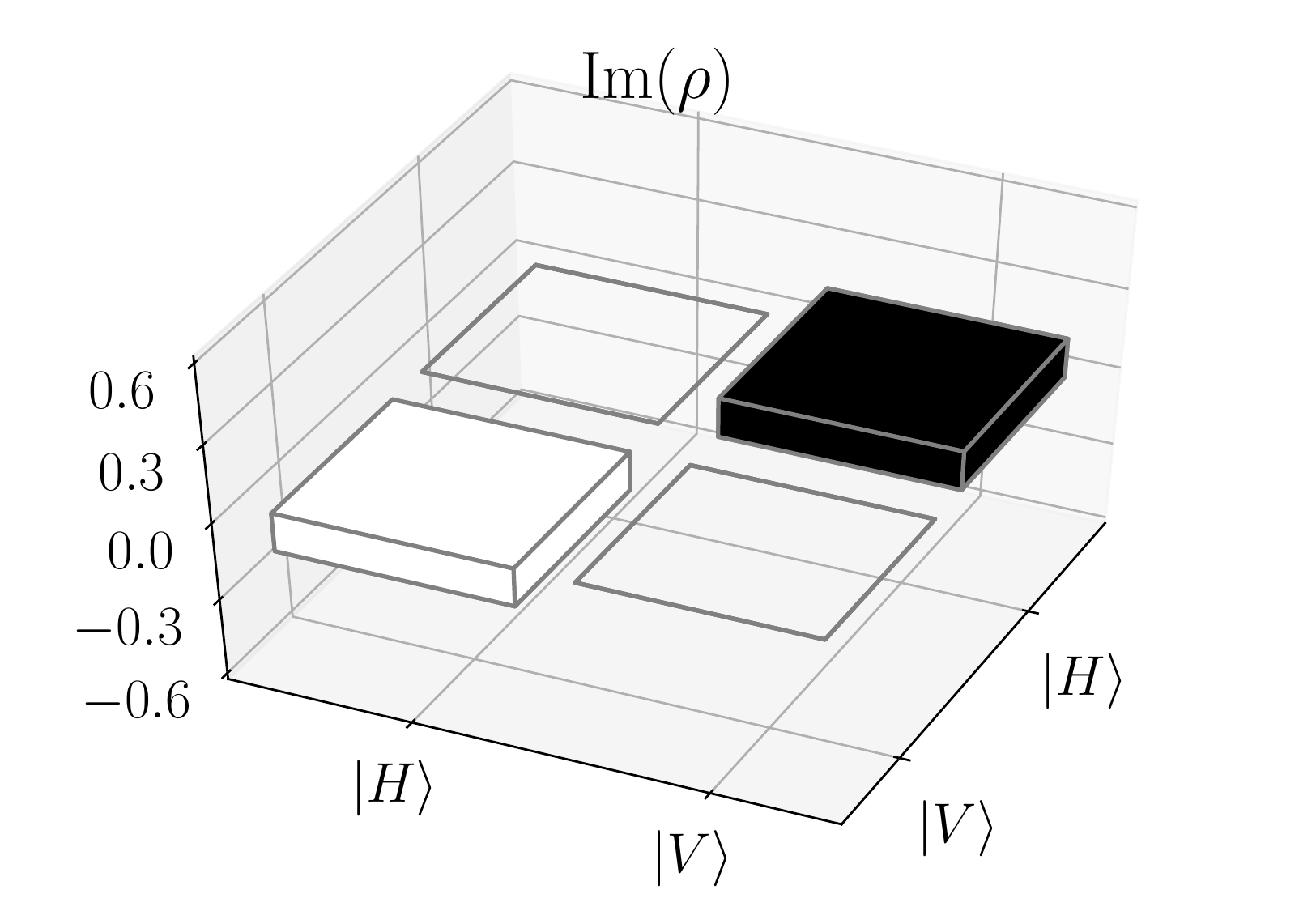} 
\caption{\label{fig:red}Real (left) and imaginary (right) parts of the density matrix for the signal photon when the photon pair is prepared in an entangled state. Black bars represent positive values, white bars negative values.}
\end{figure*}

Next, we prepare photon pairs in an entangled state close to the Bell state $\ket{\Phi^+} = \frac{1}{\sqrt{2}} \left( \ket{H}\ket{H} + \ket{V}\ket{V}\right)$, by pumping both downconversion crystals. We now consider the scenario in which bits are produced by measuring each photon in the $HV$ basis and associating the two-photon coincidence events $HH$ and $VV$ with the bits 0 and 1. (Such entangled-state measurements have been used in several QRNGs.\cite{Fiorentino:2006:lm,Fiorentino:2007:ll,Bronner:2009:ll}) If the state were indeed $\ket{\Phi^+}$, then there is maximum quantum randomness in the measurement outcomes $HH$ and $VV$, and one expects a uniform random sequence of bits. 

To assess the prepared two-photon state and the randomness it implies, we insert half- and quarter-wave plates into the signal and idler beams and tomographically reconstruct the two-photon density matrix from Eq.~\eqref{eq:dm3}. The result is 
\begin{widetext}
\begin{equation}\label{eq:1}
\rho = \begin{pmatrix} 
 0.409 &         -0.111+0.052\I & 0.009-0.148\I &
   0.360-0.182\I \\
-0.111-0.052\I  & 0.056 &    -0.003-0.006\I &
  -0.052-0.065\I \\
  0.009+0.148\I & -0.003+0.006\I &  0.030 & 
  -0.019+0.096\I \\
  0.360+0.182\I & -0.052+0.065\I & -0.019-0.096\I & 
   0.505     \end{pmatrix},  
\end{equation}
\end{widetext}
which is shown in Fig.~\ref{fig:tomo4d}. We first quantify the closeness of the reconstructed state to the Bell state $\op{\rho}_+=\ketbra{\Phi^+}{\Phi^+}$ by calculating the fidelity. We find $F(\rho, \rho_+) = 0.904$, indicating that the prepared state is well described by $\ket{\Phi^+}$. As an additional check, we confirm the presence of polarization entanglement by performing a CHSH--Bell test.\cite{Dehlinger:2002:tt} We find $S=2.457 \pm 0.002$ (the error is estimated from statistical fluctuations of the photon counts), showing a clear violation of the bound $S \le 2$ for local-realistic theories.

While we have reconstructed the $4 \times 4$ density matrix of the photon pair, the min-entropy bound~\eqref{eq:fio} is calculated from a $2\times 2$ density matrix. Therefore, to apply Eq.~\eqref{eq:fio} we must restrict the measured density matrix to the subspace relevant to the generation of bits from the $HH$ and $VV$ coincidences, which is the space spanned by $\ket{H}\ket{H}$ and $\ket{V}\ket{V}$. The portion of the density matrix associated with this subspace is $\rho_\text{sub}=\begin{pmatrix} \tilde{\rho}_{11}&\tilde{\rho}_{14}\\ \tilde{\rho}_{41} &\tilde{\rho}_{44}\end{pmatrix}$, where the $\tilde{\rho}_{ij}$ are matrix elements of the two-photon density matrix that have been renormalized such that the diagonal elements add up to one. We stress that the density matrix $\rho_\text{sub}$ is not to be confused with the reduced density matrix for a single photon obtained from a partial trace over the two-photon density operator. Rather, it is a two-photon density matrix (it describes the statistics of two-photon coincidence events) that has been limited to a two-dimensional state space defined by the coincidences of interest.

Using the data given in Eq.~\eqref{eq:1}, this subspace matrix $\rho_\text{sub}$ is 
\begin{equation}
\rho_\text{sub} = \begin{pmatrix} 
0.447 &        0.394-0.199\I \\
0.394+0.199\I & 0.553
\end{pmatrix},  
\end{equation}
shown in Fig.~\ref{fig:sub}. The probabilities for $HH$ and $VV$ are $p_{HH}=\tilde{\rho}_{11}=0.447$ and $p_{VV}=\tilde{\rho}_{22}= 0.553$, which tells us that the bit generation would be biased toward the bit 1. Such a bias would have to be removed by postprocessing. It  could  also, of course, be reduced by tuning the state, but here we purposely keep such imperfections to illustrate the issue of bias. The magnitude of the off-diagonal elements representing quantum coherence between $\ket{H}\ket{H}$ and $\ket{V}\ket{V}$ is $C=\abs{\tilde{\rho}_{14}}=\abs{\tilde{\rho}_{41}}=0.441$, which is in the vicinity of the maximum value $C_\text{max}=\sqrt{\tilde{\rho}_{11}\tilde{\rho}_{44}}=0.5$ and indicates a substantial amount of quantum randomness in the outcomes $HH$ and $VV$. Using this value $C=0.441$ in Eq.~\eqref{eq:fio} gives a minimum min-entropy of $H^\text{min}_\infty = 0.443$ per measured coincidence event $HH$ or $VV$. 

\subsection{\label{sec:meas-one-phot}Bit generation from measurements on a photon in an entangled state}

We again prepare the entangled state $\ket{\Phi^+}$ as in Sec.~\ref{sec:photon-pair-an}, but this time we consider the case where we produce the bits not from two-photon coincidence events but from $HV$ measurements on the signal photon only, as in Sec.~\ref{sec:bit-generation-from}. (Note that this is the scenario discussed in Sec.~\ref{sec:rand-mixed-stat}, where Alice measures photons entangled with Eve's photons.) The density matrix of this single-photon state (the reduced state) is obtained from the two-photon density matrix by averaging over the outputs $A$ and $A'$ of the idler (i.e., the signal events are not conditioned on the polarization of the idler, since we assume that Alice has access only to measurements on the signal photon). Formally, this averaging is represented by a partial trace over the two-photon density matrix. The resulting density matrix is 
\begin{equation} 
\rho = \begin{pmatrix} 
0.439 &         -0.130+0.148\I \\
-0.130-0.148\I & 0.561
\end{pmatrix},  
\end{equation}
which is shown in Fig.~\ref{fig:red}. The off-diagonal elements are now small: Their magnitude is only 0.197, far below the value $C_\text{max}=\sqrt{\rho_{11}\rho_{22}}=0.5$ one would have for a pure state with full coherence between $\ket{H}$ and $\ket{V}$. Accordingly, the corresponding min-entropy bound~\eqref{eq:fio} is only $H^\text{min}_\infty = 0.060$, indicating a low amount of randomness. As discussed in Sec.~\ref{sec:rand-mixed-stat}, the explanation for this loss of randomness lies in the quantum correlations between signal and idler photons, which preclude the possibility of observing coherence between $\ket{H}$ and $\ket{V}$ on the signal photon. The fact that the entangled second photon could be used to obtain information about Alice's photon forces the disappearance of coherence (and thus of guaranteed quantum randomness) on Alice's side. 

Regardless of what (if any) measurements are performed on the idler, Alice will always get a sequence of $H$s and $V$s that will appear random to her. Thus, she would not be able to detect Eve's presence based on the results of her bit-generating procedure. Fortunately, however, she can find out about the lack of guaranteed randomness from her tomographic reconstruction of the state of the signal photon. She will not know why there is a lack of randomness (she does not have knowledge of the two-photon entangled state), but the state measurement and the min-entropy value obtained from it will alert her that her bit source is not producing sufficient randomness.

\section{Conclusions}

We have considered different scenarios for the production of random bits from polarization measurements of photons, and explored, both theoretically and experimentally, how knowledge of the quantum state of the photons can help quantify the presence of randomness (defined as unpredictability). Specifically, we discussed how the amount of quantum coherence in the state between the two possible outcomes of the polarization measurement indicates the amount of fresh randomness that can be produced by the measurement. Here we also made use of a quantitative connection (proven in Refs.~\onlinecite{Fiorentino:2006:lm,Fiorentino:2007:ll}) between such coherence and a lower bound on the min-entropy of the source, which gives the most conservative estimate of the randomness of the bit-generating process. Throughout, we have emphasized a distinction between randomness resulting from a lack of information, and quantum randomness rooted in the indeterministic nature of individual quantum measurements. The min-entropy estimated from the quantum state can jointly quantify both sources of randomness. 

In our experiment, we tomographically measured the quantum state of photons prepared in nonentangled and entangled polarization states, and used this state information to calculate a lower bound on the min-entropy for different choices of the bit-generating polarization measurement. We found a large min-entropy, and hence a high amount of randomness, when the bits were produced from measurements on a photon prepared in a nonentangled superposition state, and when they were produced from joint measurements on pairs of photons prepared in an entangled state. This can be understood from the presence of a large amount of quantum randomness, coupled with near-equalized quantum probabilities for the measurement outcomes. 

By contrast, we found a low min-entropy bound when the bits were obtained from measurements on only one of the photons in an entangled pair. This loss of randomness is rooted in the quantum correlations inherent in the entangled state, leading to a decrease of quantum randomness that can be guaranteed for measurements on one photon. It can also be understood in the context of the presence of an adversary who uses measurements on the other photon in the entangled pair to learn the bit sequence, thereby compromising the privacy of the sequence. Such lack of guaranteed randomness and privacy does not manifest itself in the bit sequence, but it can be detected by a measurement of the quantum state.

\vspace{.5cm}

\begin{acknowledgments}
This work was supported by the  M.~J. Murdock Charitable Trust and by the SURE program of the University of Portland.
\end{acknowledgments}


\end{document}